\documentclass[aps, preprintnumbers,amsmath,amssymb,superscriptaddress, floatfix]{revtex4}
\usepackage{graphicx, bm}
\usepackage{latexsym}
\usepackage{amsmath}
\usepackage{graphics}
\usepackage{amssymb}
\usepackage{layout}
\usepackage{verbatim}
\usepackage{amsfonts,epsfig}
\newcommand{\beq}{\begin{equation}}
\newcommand{\eeq}{\end{equation}}

\DeclareMathOperator{\trace}{tr}
\def\openone{\leavevmode\hbox{\small1\kern-0.8ex\normalsize1}}

\newcommand* {\vekc}[1]{\ensuremath{\bm{\mathcal{#1}}}}

\newcommand {\bra} [1] {\langle #1 |}
\newcommand {\ket} [1] {| #1 \rangle}
\newcommand {\bkt} [1] {\langle #1 \rangle}

\newcommand {\tbkt} [3] {\langle #1 | #2 | #3 \rangle}
\newcommand {\pd} [2] {\frac{\partial #1}{\partial #2}}
\newcommand {\td} [2] {\frac{d #1}{d #2}}

\begin{document}

\title{Steady-state spin densities and currents}
\author{Dimitrie Culcer}
\address{Condensed Matter Theory Center, Department of Physics, University of Maryland, College Park MD 20742.}
\address{Advanced Photon Source, Argonne National Laboratory, Argonne, IL 60439.} 
\address{Northern Illinois University, De Kalb, IL 60115.}
\date{\today}

\begin{abstract}
This article reviews steady-state spin densities and spin currents in materials with strong spin-orbit interactions. These phenomena are intimately related to spin precession due to spin-orbit coupling which has no equivalent in the steady state of charge distributions. The focus will be initially on effects originating from the band structure. In this case spin densities arise in an electric field because a component of each spin is conserved during precession. Spin currents arise because a component of each spin is continually precessing. These two phenomena are due to independent contributions to the steady-state density matrix, and scattering between the conserved and precessing spin distributions has important consequences for spin dynamics and spin-related effects in general. In the latter part of the article extrinsic effects such as skew scattering and side jump will be discussed, and it will be shown that these effects are also modified considerably by spin precession. Theoretical and experimental progress in all areas will be reviewed. 
\end{abstract}
\maketitle

\section{Introduction}

Spin electronics seeks to harness the spin degree of freedom of the electron, aiming to generate and maintain a spin polarization for possible use in information storage through the manufacture of magnetic memory devices. For spin-based quantum computing the manipulation of a spin polarization is also crucial. A spin polarization can be genegrated by optical, magnetic and electrical means. Optically, one uses light to transfer angular momentum to charge carriers during interband transitions.\cite{Meyer} This reliable procedure has been tested successfully decades ago, yet devices that require an input of light are currently impractical. Magnetically, one uses the Zeeman effect, to which similar observations apply. Electrically, one drives a current from a spin-polarized material into an unpolarized one and thus carry spin-polarized electrons across the interface. This works in the case of metals but modern technology relies overwhelmingly on semiconductors, and spin injection from ferromagnetic metals into semiconductors is hampered by the resistivity mismatch \cite{Schmidt} at the interface between the metal and the semiconductor, which causes most of the spin polarization to be lost at the interface. Another avenue explores the use of ferromagnetic semiconductors\cite{Jungwirth} as spin injection devices, but room temperature ferromagnetism in semiconductors remains a long-term goal. 

As a result of these challenges, purely electrical means of spin generation have been sought. The past decade has witnessed an explosion of interest in electrically-induced spin phenomena in semiconductors and metals, which has accompanied significant experimental and theoretical progress. It has been found that the application of an electric field gives rise to a nonequilibrium spin density in the bulk of the sample, discussed in Refs.\ [\cite{ivc78}-\cite{sil04}], and a nonequilibrium spin current, discussed in Refs.\ [\cite{dya71}-\cite{DyakMagRes}]. The appearance of a steady-state spin density in an electric field was predicted several decades ago \cite{ivc78} and subsequently observed in tellurium. \cite{vor79} This work was followed by a number of theoretical \cite{lev85,ede90,aro91,cha01} and experimental \cite{kato04b,GanichevJMMM,sil04} studies.
Spin currents also have a long history, their theoretical prediction stretching back almost four decades to the work of Dyakonov.\cite{dya71} The theoretical work was, however, not followed by experiments, and the topic was silent for over a quarter of a century, until the theory picked up again, led by Hirsch \cite{hir99}, Murakami \textit{et al.}\cite{mur03} and Sinova \textit{et al}.\cite{sin04} During the recent surge of interest in electrically-induced spin phenomena the focus has been principally on spin currents, with the ultimate goal of \textit{measuring} spin flow in a given direction. In the absence of a \textit{spin-ammeter}, initially the only way of detecting a spin current relied on measuring the spin accumulation it produced at the edge of the sample. This procedure, as will be discussed below, is complicated by the fact that the spin current is often not well defined and its relationship to spin accumulation is not clear. Thus one can detect a spin accumulation at the edge of the sample as a result of the spin current, but inferring the size of the spin current from such a measurement is beyond current understanding.  Nevertheless, in just a few years experiment has been surging ahead. First, in a recent experiment, Crooker and Smith \cite{cro05} imaged the flow of spin in semiconductors in the presence of electric and magnetic fields, as well as strain. Following that, spin accumulations resulting from spin currents were measured by several groups. \cite{wun05,kato04,sih05,ste06,AwschGen,AwschDriftDiff,Chang}  In what is surely a new generation of experiments, Valenzuela and Tinkham\cite{val06} made use of the fact that a spin current gives rise to a transverse charge current. Relying on a nonlocal technique, their group successfully detected a charge current flowing as a result of a spin current in Al metal. Cui \textit{et al.}\cite{cui06} also observed an electrical current induced by a spin current, though in that case the spin current was generated optically. This technique was used by several groups \cite{SaitohInvSHE,Kimura,Vila,Seki,Singapore} shortly afterwards in the study of platinum- and gold-based structures. In gold an unusually large spin current was observed\cite{Seki} even at room temperature. Room temperature spin currents were also reported by Stern \textit{et al} \cite{ste06} in ZnSe. Thus, from their initial observation four years ago, spin currents have begun an experimental revolution, and the field has matured to the point where experiment and theory interact constructively.

The phenomena discussed in this review are all related to spin-orbit interactions, which are present in the band structure and in potentials due to impurity distributions. Band structure spin-orbit interactions are frequently the most important factor determining spin dynamics in solids, and are particularly important in semiconductors, where carriers involved in steady-state processes have wave vectors well within the first Brillouin zone. At these wave vectors, while the spin-orbit contribution to the carrier energy is still smaller than the kinetic energy, it plays an important role in determining the energy spectrum. Band structure spin-orbit coupling may arise from the inversion asymmetry of the underlying crystal lattice \cite{dre55}, from the inversion asymmetry of the confining potential in two dimensions \cite{ras84}, and may be present also in inversion symmetric systems. \cite{lut56} Spin-orbit coupling is also in principle present in impurity potentials. Spin-orbit interactions in the impurity potentials cause skew scattering, or asymmetric scattering of up and down spins, giving scattering-dependent or \textit{extrinsic} contributions to spin currents. Finallly, spin-orbit interactions cause a modification of the position and velocity operators, which affects the interaction with an electric field as well as the scattering term. This last mechanism is referred to as side jump and is typically classified as extrinsic although its contributions are manifold. These extrinsic mechanisms were discovered in the anomalous Hall effect \cite{Lutt,Smit,Berger,Nozieres} but have acquired a new relevance in recent years in electrically-induced \textit{spin} phenomena.

In charge transport, the steady state is characterized by a nonequilibrium density matrix that is divergent in the clean limit, indicating a competition between the electric field, accelerating charge carriers, and scattering, which inhibits their forward motion. The density matrix is a scalar and the correction to it caused by the electric field is $\propto n_i^{-1}$, where $n_i$ is the impurity density. When considering the steady state of systems with spin-orbit interactions three different situations are distinguished: the case when only extrinsic mechanisms are important, the case when only band-structure mechanisms are important, and the case when both extrinsic and band-structure mechanisms are important. The differences between these cases are significant and nontrivial.

If band structure spin-orbit interactions are negligible and only extrinsic mechanisms are important, there is no spin precession and the correction to the spin density matrix due to an electric field is very similar to the correction to the scalar density matrix in charge transport. This correction is $\propto n_i^{-1}$ and it gives rise to a steady-state spin current but not to a steady-state spin density. There have been many studies of extrinsic spin currents in semiconductors (Refs.~[\cite{eng05}-\cite{EwelinaC}]) and metals (Refs.~[\cite{hir99},\cite{Shche}].)

Nonequilibrium corrections that arise as a result of band structure spin-orbit coupling in crystal Hamiltonians represent a different kind of interplay between the electric field and scattering processes. The spin-orbit splitting of the bands gives rise to spin-dependent scattering even from spin-independent potentials, and spin currents and spin densities in an electric field arise from linearly independent contributions to the density matrix. The steady-state density matrix contains a contribution due to precessing spins and one due to conserved spins. Steady-state corrections $\propto n_i^{-1}$ are associated with the \emph{absence} of spin precession and steady-state corrections independent of $n_i$ are associated with spin precession. Steady-state corrections $\propto n_i^{-1}$ give rise to spin densities in external fields while steady-state corrections independent of $n_i$ give rise to spin currents in external fields. Scattering between these two distributions induces significant corrections to steady-state spin currents, and may cause spin currents to vanish.

The phrase \emph{spin current} refers to the flow of spins across a sample and, if spin were conserved, one could distinguish between spin-up and spin-down charge currents. In systems with band structure spin-orbit interactions the spin current is not well defined \cite{shi06,sug06,Bryksin} and its relationship to spin accumulation is unclear. Spin transport in these systems usually does not involve charge transport as the charge currents in the direction of spin flow cancel out. In addition, since spin currents may be accompanied by steady-state spin densities, this makes it more difficult to separate experimentally the signals due to these contributions. The study of spin currents in systems with band structure spin-orbit interactions has thus encountered a number of profound physical issues on which no consensus exists at present. There has been considerable research has on definitions of spin currents in such systems (Refs.~[\cite{shi06}-\cite{Tokatly}]), the role played by the symmetry of the underlying crystal lattice,\cite{dim07} whether spin currents as a result of band structure spin-orbit coupling are transport or background currents, \cite{ras04} the relationship between band structure spin currents and spin accumulation, (Refs.~[\cite{TserkovnyakAcc}-\cite{ada05}]) and the form of the Maxwell equations in systems with band structure spin-orbit coupling.\cite{wang06,BernevigMaxwell} Most theoretical studies have focused on metals, common semiconductors and asymmetric quantum wells with band structure spin-orbit interactions, such as Refs.~[\cite{sin04} -\cite{GuoPt}].

When both band structure and extrinsic mechanisms are present, the situation becomes considerably more complicated. There is no simple association between spin conservation and spin densities, or between spin precession and spin currents. Both spin densities and currents may contain terms $\propto n_i^{-1}$ and independent of $n_i$. A careful analysis shows that the presence of band structure spin-orbit interactions causes skew scattering and side jump to behave very differently in the steady state, and under certain circumstances, the contributions of these extrinsic mechanisms to the spin current vanish. The interplay of band structure and extrinsic mechanisms has been considered in Refs.~[\cite{EwelinaP}-\cite{ChengWu}].

This review will cover steady-state spin densities and currents within the framework of a density matrix theory. Such a framework is important in order to demonstrate the unity behind all observed phenomena. I will consider large, uniform systems, and work in momentum space. Although many observations in this entry are general, the discussion will focus on non-interacting spin-1/2 electron systems, which are pedagogically easier. This review is based on a kinetic-equation  formalism equivalent to linear response theory and correspondences with other methods will be identified, such as linear response theories based on Green's functions and semiclassical wave packet dynamics.

This review article is structured as follows. Section 2 will introduce a density matrix picture of spin dynamics, briefly sketching the way different theories stem from the quantum Liouville equation. In section 3 a brief introduction to the spin-orbit interaction in spin-1/2 systems is given, after which spin densities and spin currents as a result of band structure spin-orbit interactions are discussed, as well as their relationship to spin precession. Section 4 is devoted to spin densities and spin currents that arise as a result of extrinsic mechanisms, together with the case when both extrinsic and band-structure mechanisms are present. Section 5 discusses briefly various definitions of spin currents and the relationship between spin currents and spin acumulation. Section 6 outlines the role of the underlying crystal lattice in the establishment of spin densities and currents and section 5 is concerned with open issues such as the definition and nature of spin currents, as well as the relationship between spin currents and spin accumulation. In section 7 the experimental situation is summarized and in closing future directions are discussed. Related topics such as the quantum spin-Hall effect \cite{QSHE} and the spin-Hall insulator \cite{SHI1} are beyond the scope of this review.

\section{Density matrix picture of spin dynamics}

Electrically-induced spin phenomena encompass a wide variety of processes that have been studied using different methods. In order to bring out the unity behind the processes involved, as well as behind the theoretical approaches employed, it will be useful to have a unified framework in which electrically-induced spin phenomena can be discussed. In this section such a framework will be constructed beginning with the quantum Liouville equation. This equation, in one form or another, is the starting point of most theories of spin dynamics, and this section will outline the way various approaches are related to each other. The focus will be on systems with large homogeneous systems with long mean free paths, and diffusion will not be considered explicitly. Electric fields will be assumed uniform. 

\subsection{Quantum Liouville equation}

A system of non-interacting spin-1/2 electrons is represented by a one-particle density operator $\hat \rho$. The expectation value of an observable represented by a Hermitian operator $\hat O$ is given by Tr$(\hat\rho\hat O)$, where Tr denotes the most general operator trace, which involves summation over discrete degrees of freedom and integration over continuous ones. The usual matrix trace will be denoted by $\trace$. The dynamics of $\hat \rho$ are described by the quantum Liouville equation,
\begin{equation}\label{Liouville}
\td{\hat\rho}{t} + \frac{i}{\hbar} \, [\hat H + \hat{U}^{dis} + e{\bm E}\cdot\hat{\bm r}, \hat \rho] = 0.
\end{equation}
The Hamiltonian $\hat H$ contains contributions due to the kinetic energy and spin-orbit coupling, while $\hat{\bm r}$ is the position operator. The effect of the lattice-periodic potential of the ions is taken into account through a replacement of the carrier mass by the effective mass. The potential $\hat{U}^{dis}$ accounts for scattering processes, which may be due to impurities, phonons, surface roughness, or other perturbations. This review focuses on impurity scattering, as the effects discussed are frequently observed at low temperatures, where scattering due to phonons may be neglected. It is assumed that the electric field is small and a solution is sought to first order in the electric field. The solution of the Liouville equation (\ref{Liouville}) in the absence of the external field is provided by $\hat{\rho}_0$, the Fermi-Dirac function. The correction due to the external field, $\hat{\rho}_E$, satisfies
\begin{equation}\label{rhoE}
\td{\hat{\rho}_E}{t} + \frac{i}{\hbar} \, [\hat{H} + \hat{U}^{dis}, \hat{\rho}_E] = - \frac{i}{\hbar} \, [e{\bm E}\cdot\hat{\bm r}, \hat{\rho}_0].
\end{equation}
Equation (\ref{rhoE}) is projected onto a complete set of time-independent states of definite wave vector $\{ \ket{{\bm k}s} \}$. These states are not assumed to be eigenstates of $\hat H$. The matrix elements of $\hat \rho$ in this basis will be written as $\rho_{{\bm k}{\bm k}'} \equiv \rho^{ss'}_{{\bm k}{\bm k}'} = \bra{{\bm k}s} \hat\rho \ket{{\bm k}'s'}$, with corresponding notations for the matrix elements of $\hat H$ and $\hat{U}^{dis}$. Spin indices will not be shown explicitly in the subsequent discussion, and it will be understood that the quantities $\rho_{{\bm k}{\bm k}'}$, $H_{{\bm k}{\bm k}'}$, and $U^{dis}_{{\bm k}{\bm k}'}$ are matrices in spin space. $\rho_{{\bm k}{\bm k}'}$ is referred to as the density matrix. With our choice of basis functions of definite wave vector, matrix elements of the Hamiltonian $H_{{\bm k}{\bm k}'} = H_{{\bm k}} \, \delta_{{\bm k}{\bm k}'}$ are diagonal in ${\bm k}$. However, since the Hamiltonian contains spin-orbit coupling terms, matrix elements $H_{{\bm k}}$ are generally off-diagonal in spin space.
Matrix elements of the scattering potential $U^{dis}_{{\bm k}{\bm k}'}$ are off-diagonal in ${\bm k}$. Matrix elements diagonal in ${\bm k}$ in the scattering potential would lead to a redefinition of $H_{{\bm k}}$, which is analogous, in Green's function formalisms, to the offset introduced by the real part of the self energy. Scattering is assumed elastic and impurities uncorrelated, and the normalization is such that the configurational average of $\bra{{\bm k}s}\hat{U}^{dis}\ket{{\bm k}'s'}\bra{{\bm k}'s'}\hat{U}^{dis}\ket{{\bm k}s}$ is $(n_i |U_{{\bm k}{\bm k}'}|^2 \delta_{ss'})/V$, where $n_i$ is the impurity density, $V$ the crystal volume and $U_{{\bm k}{\bm k}'}$ the matrix element of the potential of a single impurity. Configurational averages over terms of higher order in $\hat{U}^{dis}$ are performed in a similar fashion. \cite{Rammer} It is assumed that $\varepsilon_F\tau/\hbar \gg 1$, where $\varepsilon_F$ is the Fermi energy and $\tau$ a characteristic scattering time. This is equivalent to the assumption that the carrier mean free path is much larger that the de Broglie wavelength. None of the methods presented below are valid once $\varepsilon_F\tau_p/\hbar$ becomes comparable to unity.

Spin densities and spin currents are obtained as expectation values of spin and spin current operators. The spin operator is given by $s^\sigma = (\hbar/2)\, \sigma^{\sigma}$, where $\sigma^\sigma$ is a Pauli spin matrix. The spin current operator will be taken to be $\hat{\mathcal{J}}^\sigma_{i} = (1/2)\, \{ s^\sigma, v^i \}$, where the velocity operator is $v^i = (1/\hbar) \, \partial H_{\bm k}/\partial k_i$. Both operators are diagonal in ${\bm k}$. One is therefore primarily interested in the part of the density matrix diagonal in ${\bm k}$. With this observation in mind, $\rho_{{\bm k}{\bm k}'}$ is divided into a part diagonal in ${\bm k}$ and a part off-diagonal in ${\bm k}$ as $\rho_{{\bm k}{\bm k}'} = f_{{\bm k}} \, \delta_{{\bm k}{\bm k}'} + g_{{\bm k}{\bm k}'}$, where, in $g_{{\bm k}{\bm k}'}$, it is understood that ${\bm k} \ne {\bm k}'$. Further, $f_{\bm k}$ is decomposed into a scalar part and a spin-dependent part, $f_{\bm k} = n_{\bm k} \,\openone + S_{\bm k}$, with $\openone$ the identity matrix. In an electric field, $f_{\bm k} = f_{0{\bm k}} + f_{E{\bm k}}$, where $f_{E{\bm k}}$ is a correction linear in ${\bm E}$ and has a corresponding decomposition into a scalar part and a spin-dependent part. At the end one takes the trace of the spin and spin current operators with $f_{E{\bm k}}$ given by Eq.\ (\ref{fGF}). The final result is usually expressed in many ways. 

It should be noted that the electric field induces coherence between bands. In equilibrium the Fermi-Dirac function $f_{0{\bm k}}$ commutes with the Hamiltonian and is stationary. However the source term due to ${\bm E}$ in Eq.\ (\ref{rhoE}) does not in general commute with the Hamiltonian. If the problem is considered in the basis of eigenstates of $H_{\bm k}$, the source term couples different energy bands. If a basis is used in which one spin component is a good quantum number, then the source term couples up and down spins. 

\subsection{Linear response Green's function formalism}

The linear response Green's function formalism is closely related to the density matrix formalism but relies on a somewhat different way of regarding the problem \cite{Rammer}. The electric field-induced correction to the density matrix $\hat{\rho}_E$ is found by applying the time evolution operator corresponding to the \textit{total} Hamiltonian $\hat{H} + \hat{U}^{dis}$, incuding disorder, in Eq.\ (\ref{rhoE}). The matrix elements of the time evolution operator in the basis $\{ \ket{{\bm k}s} \}$ constitute the Green's function $G_{{\bm k}{\bm k}'}^{ss'} (t)$, defined by 
\begin{equation}
G_{{\bm k}{\bm k}'}^{ss'} (t) = \tbkt{{\bm k}s}{e^{-i(\hat{H} + \hat{U}^{dis})t/\hbar}}{{\bm k}'s'}.
\end{equation}
Spin indices are again suppressed. All the information about the system is contained in the Green's function, which is also referred to as the propagator, since it gives the amplitude for a particle to propagate in time $t$ from state $\ket{{\bm k}s}$ to state $\ket{{\bm k}'s'}$. The Green's function is also the kernel of the Schrodinger equation and represents the response of the system to a perturbation that is localized in real space or momentum space. The total perturbation is built up as a sum of localized ones. The problem can be formulated in any basis but for our purposes it is still the most convenient to work in the basis of eigenstates of $H_{so{\bm k}}$. In general the Green's function is a matrix in spin space. For practical purposes one also defines the retarded and advanced Green's functions, $G^R$ and $G^A$, as $G^R = -i \, G \, \theta(t)$ and $G^A = i\, G \, \theta(-t)$, with $\theta$ the Heaviside step function. The electric field-induced correction to the part of the density matrix diagonal in ${\bm k}$ is
\begin{equation}\label{fGF}
f_{E{\bm k}} = - \frac{ie{\bm E}}{\hbar} \cdot \sum_{{\bm k}'} \int_{-\infty}^\infty dt' \, G_{{\bm k}{\bm k}'}^R (t') [\hat{\bm r}, \hat{\rho}_0]_{{\bm k}'} G_{{\bm k}'{\bm k}}^A (-t')
\end{equation}
At this stage it is easiest to carry out a Fourier transformation with respect to time. The integration in Eq.\ (\ref{fGF}) has the same form except the time variable is replaced by the frequency variable $\omega$
\begin{equation}\label{fGFw}
f_{E{\bm k}} = - \frac{ie{\bm E}}{\hbar} \cdot \sum_{{\bm k}'} \int_{-\infty}^\infty d\omega \, G_{{\bm k}{\bm k}'}^R (\omega) [\hat{\bm r}, \hat{\rho}_0]_{{\bm k}'} G_{{\bm k}'{\bm k}}^A (-\omega).
\end{equation}
The correction $f_{E{\bm k}}$ to the density matrix needs to be averaged over impurities, and it is assumed henceforth that only the impurity averaged density matrix is of interest. We denote this impurity average by the symbol $\bkt{}$, but the impurity-averaged density matrix will be abbreviated as$\bkt{f_{E{\bm k}}} \equiv f_{E{\bm k}}$. The full Green's function is determined by the total Hamiltonian $\hat{H} + \hat{U}^{dis}$ including the scattering potential. The Green's function is expanded in the scattering potential, and one is only interested in the Green's function \textit{averaged} over impurity configurations. For uncorrelated impurities the impurity-averaged Green's function, denoted by $\bkt{G}$, obeys the recursion relation
\begin{equation}
\bkt{G^{R/A}({\bm k}, \omega)} = G^{R/A}_0({\bm k}, \omega) + G^{R/A}_0({\bm k}, \omega) \Sigma^{R/A}({\bm k}, \omega) G^{R/A}({\bm k}, \omega) 
\end{equation}
in which $G_0$ is the equilibrium Green's function, corresponding to the Hamiltonian $\hat{H}$ without disorder, and the self energy $\Sigma$ is typically evaluated in the first Born approximation
\begin{equation}
 \Sigma^{R/A}({\bm k}, \omega) = n_i \int\frac{d^d k'}{(2\pi)^d}\, |U_{{\bm k}{\bm k}'}|^2\, G_0^{R/A}({\bm k}', \omega) 
\end{equation}
with $d$ the dimensionality of the system. The correction $f_{E{\bm k}}$ to the density matrix depends on the product of two Green's functions. After the average over impurity configurations we obtain terms which are expressible as a product of two individual impurity-averaged Green's functions and cross terms, which connect different Green's functions, and which are not expressible as a product. These form a vertex $ \Gamma_{{\bm k}{\bm k}'}$ defined by 
\begin{equation}
\bkt{G^R({\bm k}, \omega) G^A({\bm k}', \omega)} = \bkt{G^R({\bm k}, \omega)} \bkt{G^A({\bm k}, \omega)} + \Gamma_{{\bm k}{\bm k}'}(\omega)\, \bkt{G^R({\bm k}', \omega)} \bkt{G^A({\bm k}', \omega)}. 
\end{equation}
The vertex is in general expressed as a sum of different classes of diagrams, of which the most relevant to transport are the ladder diagrams and the maximally crossed diagrams. It also satisfies the recursion relation
\begin{equation}
 \Gamma_{{\bm k}{\bm k}'}(\omega) =  M_{{\bm k}{\bm k}'}(\omega) + \int \frac{d^dk''}{(2\pi)^d}\, M_{{\bm k}{\bm k}''}(\omega) \bkt{G^R({\bm k}'', \omega)} \bkt{G^A({\bm k}'', \omega) \Gamma_{{\bm k}''{\bm k}'}(\omega)}
\end{equation}
where $M_{{\bm k}{\bm k}'}(\omega)$ can also be expressed in terms of diagrams as shown in Ch.~8 of Ref.~[\cite{Rammer}]. Once the impurity-averaged Green's function $\bkt{G}$ and the vertex are known, $f_{E{\bm k}}$ can be found. It will be noticed that in the linear response Green's function approach disorder is contained in the self energy $ \Sigma({\bm k})$ and in the vertex $ \Gamma_{{\bm k}{\bm k}'}$. The response function contains all the disorder up to the desired approximation. Once the Green's function and the vertex function are found, the trace with the spin and spin current operators is taken.

\subsection{Kinetic equation formalism}

The kinetic equation formalism can be derived from the quantum Liouville equation, as will be done below, or using a Keldysh Green's function formalism.\cite{shy06} The kinetic equation obtained is naturally the same. Starting from the quantum Liouville equation, this can be broken down into equations for $f_{{\bm k}}$ and $g_{{\bm k}{\bm k}'}$
\begin{subequations}
\begin{eqnarray}
\td{f_{E{\bm k}}}{t} + \frac{i}{\hbar} \, [H_{{\bm k}}, f_{E{\bm k}}] & = & - \frac{i}{\hbar} \, [\hat U, \hat{g}_E]_{{\bm k}{\bm k}} - \frac{i}{\hbar} \, [e{\bm E}\cdot\hat{\bm r}, \hat{\rho}_0]_{{\bm k}{\bm k}}, \\ [1ex] 
\td{g_{E{\bm k}{\bm k}'}}{t} + \frac{i}{\hbar} \, [\hat H, \hat{g}_E]_{{\bm k}{\bm k}'} & = & - \frac{i}{\hbar} \, [\hat U, \hat{f}_E + \hat{g}_E]_{{\bm k}{\bm k}'} \label{eq:g}.
\end{eqnarray}
\end{subequations}
In the first Born approximation the solution to Eq.\ (\ref{eq:g}) can be written as
\begin{equation}
g_{E{\bm k}{\bm k}'} = - \frac{i}{\hbar} \, \int_0^\infty dt'\, e^{- i \hat H t'/\hbar} \left[\hat U, \hat{f}_E (t - t') \right] e^{i \hat H t'/\hbar}|_{{\bm k}{\bm k}'}.
\end{equation}
Since $\varepsilon_F\tau_p/\hbar \gg 1$, we shall expand $\hat f(t - t')$ in the time integral around $t$ and, noting that terms beyond $\hat f(t)$ are of higher order in the scattering potential, we shall only retain the first term, $\hat f(t)$. The equation for $f_{{\bm k}}$ then becomes
\begin{subequations}\label{eq:FermiJfk}
\begin{equation}\label{eq:Fermi}
\td{f_{E{\bm k}}}{t} + \frac{i}{\hbar} \, [H_{{\bm k}}, f_{E{\bm k}}] + \hat J (f_{E{\bm k}}) = \frac{e{\bm E}}{\hbar} \cdot \bigg( \pd{f_{0{\bm k}}}{{\bm k}} - i [ \bm{\mathcal R}, f_{0{\bm k}} ] \bigg) \end{equation} 
in which the scattering term $\hat J (f_{{\bm k}})$ is given by
\begin{equation}
\label{eq:Jfk} \hat J (f_{E{\bm k}}) = \frac{1}{\hbar^2} \,\int_{0}^\infty dt'\, \left[\hat U, e^{- i \hat H t'/\hbar} \left[\hat U, \hat{f}_E (t) \right] e^{ i \hat H t'/\hbar} \right]_{{\bm k}{\bm k}}
\end{equation}
\end{subequations}
and the source term on the RHS contains the \textit{covariant} derivative with respect to ${\bm k}$, which takes into account the fact that the basis functions themselves may depend on ${\bm k}$. It includes the gauge connection matrix $\bm{\mathcal R} \equiv \bm{\mathcal R}_{ss'} = \tbkt{{\bm k}s}{i\partial/\partial {\bm k}}{{\bm k}s'}$. The response function, as well as the scattering term need to be averaged over impurity configurations, and the notation $f_{E{\bm k}}$ is used henceforth as above to denote the impurity-averaged density matrix. The scattering term can be expanded further in the basis in spin space spanned by spin eigenstates $\ket{\uparrow}$ and $\ket{\downarrow}$ (the Pauli basis.) In this basis the connection matrix $\bm{\mathcal R}$ vanishes and the scattering potential $U_{{\bm k}{\bm k}'} = \mathcal{U}_{{\bm k}{\bm k}'} \openone$ is diagonal in spin space. Converting sums over wave vector into integrals $\sum_{{\bm k}'} \rightarrow V \int \frac{d^dk'}{(2\pi)^d}$, and performing the time integral the scattering term can be expressed in the form $(\hat{J}_0 + \hat{J}_b)\, (n_{\bm k}) + \hat{J}_0(S_{\bm k})$, with
\begin{subequations}
\label{eq:scattering}
\begin{eqnarray}
\hat{J}_0 \, (S_{\bm k}) & = & \frac{2\pi n_i}{\hbar} \int \frac{d^dk'}{(2\pi)^d}\, |\mathcal{U}_{{\bm k}{\bm k}'}|^2 (S_{\bm k} - S_{{\bm k}'}) \delta(\varepsilon_{0{\bm k}} - \varepsilon_{0{\bm k}'}) \hspace{2em} \\ [1ex] 
\label{eq:scatteringb} \hat{J}_b \, (n_{\bm k}) & = & \frac{2\pi n_i}{\hbar} \int \frac{d^dk'}{(2\pi)^d}\, |\mathcal{U}_{{\bm k}{\bm k}'}|^2(n_{\bm k} - n_{{\bm k}'})\, \frac{1}{2} \, {\bm \sigma}\cdot({\bm \Omega}_{{\bm k}} - {\bm \Omega}_{{\bm k}'}) \pd{}{\varepsilon_{0{\bm k}}} \delta(\varepsilon_{0{\bm k}} - \varepsilon_{0{\bm k}'}) \nonumber.
\end{eqnarray}
\end{subequations}
The term $\hat{J}_b \, (n_{\bm k})$ in Eq.\ (\ref{eq:scatteringb}) illustrates the fact that, when spin-orbit interactions are present in the band structure, even spin-independent scattering potentials give rise to spin-dependent terms in the scattering integral. For potentials diagonal in spin space and spin-degenerate bands, Eq.\ (\ref{eq:Jfk}) simplifies to the customary Fermi's golden rule. Therefore, Eq.\ (\ref{eq:FermiJfk}) is a generalization of Fermi's golden rule that explicitly takes into account the spin degree of freedom.

\subsection{Boltzmann-wave packet formalism}

The density matrix is the most complete description of a physical system. It contains all the relevant physics, including interband coherence due to the electric field and scattering. It represents the system as a whole and accounts for the individual particle motion, which occurs between collisions, as well as for the distribution of electrons in phase space and the way it is altered by collisions. As was shown above, linear response theories based on Green's functions and theories based on kinetic equations originate directly from the Liouville equation for the density matrix. A third approach, which may be called a semiclassical or Boltzmann-wave packet approach, separates the motion of the charge carriers from their distribution in phase space. To determine the dynamics of the charge carriers one envisages them as being represented by wave packets, and the phase space distribution of wave packets is given by a function of the Boltzmann form. One use of the word \emph{semiclassical}, which will be adopted in this work, is to refer to theories which consider the position and momentum of a particle simultaneously. Semiclassical approaches exploit the smooth variation of transport fields on atomic length scales in order to provide intuitive descriptions of steady-state processes, which can be easily extended to cover inhomogeneous systems and spatially dependent fields.

The semiclassical approach is also closely related to the density matrix and the Liouville equation. Firstly, in order to separate the particle dynamics and the phase space distribution one must be able to label the particles by a band index, therefore the theory must be formulated in the basis of eigenstates of $H_{\bm k}$. Since these eigenstates are usually ${\bm k}$-dependent it is the covariant derivative that enters the source term in the kinetic equation. The point at which semiclassical theory branches out of the kinetic equation is Eq.\ (\ref{eq:Fermi}). In order to turn the density matrix into a Boltzmann distribution function one takes the diagonal elements of this equation. If we consider for simplicity a spin-1/2 system, which is made up of two bands that will be called 1 and 2, we can label the diagonal elements of the density matrix as $f_{1{\bm k}}$ and $f_{2{\bm k}}$. These become the Boltzmann distributions for bands 1 and 2. The commutator $[H_{\bm k}, f_{\bm k}]$ has no diagonal elements and $\hat{J}(f_{\bm k})$ reduces to one scattering term for each band, which is given by Fermi's Golden Rule, plus interband scattering terms which couple $f_{1{\bm k}}$ and $f_{2{\bm k}}$. In this way Eq.\ (\ref{eq:Fermi}) reduces to two coupled Boltzmann equations for the two bands and the phase space distribution has been separated out of the kinetic equation.

In reducing the kinetic equation to a series of Boltzmann equations one neglects interband coherence due to the electric field, and this must be recovered elsewhere. In the semiclassical approach, this occurs in several ways. Firstly, one needs to consider the nature of the carriers in band $s$ and invoke explicitly the fact that carriers are described by wave packets. This is the point at which the formalism turns into a true semiclassical theory. The construction of a wave-packet representing a charge and spin carrier in band $s$, which has real and $k$-space coordinates $({\bm r}_{cs}, {\bm k}_{cs})$, has been thoroughly treated by Sundaram and Niu \cite{Sundaram} and its extension to multiple bands was carried out in Ref.\ [\cite{dim05a}]. The macroscopic distribution associated with a quantum mechanical operator $\hat{O}$ is no longer given simply by the expectation value of the operator $\hat{O}$, but involves explicitly a sum over wave packets centered at $({\bm r}_{cs}, {\bm k}_{cs})$. For example the spin density distribution is 
\begin{equation}\label{Swpk}
S^\sigma({\bf r}, t) = \sum_s \int d^3k_{cs} \int d^3r_{cs} f_s({\bm k}_{cs}, t)\bkt{\delta({\bf r} - {\bf \hat r})\hat{s}^\sigma}_s,
\end{equation}
where the bracket indicates quantum mechanical average over the wavepacket with charge centroid $({\bf r}_{cs}, {\bf k}_{cs})$. An analogous expression exists for the spin current distribution. The wave-packet expectation values $\bkt{}_s$ in Eq.\ (\ref{Swpk}) eventually yield functions of  $({\bf r}_{cs}, {\bf k}_{cs})$, the dynamics of which are discussed below. It may appear contradictory that one has to integrate over the real-space coordinates of the wavepackets ${\bm r}_{cs}$ even in the case of homogeneous systems studied in this work, where the distribution function is not a function of ${\bm r}_{cs}$. This is because even in a homogeneous system of the center of charge and the center of spin are not the same, so variations of the spin distribution on the spatial scales comparable to that of the wave packet need to be taken into account, as shown in Fig.\ 1. All these terms must be considered in order to reach agreement with approaches based directly on the Liouville equation.

In a constant uniform electric field ${\bm E}$ the coordinates of the wave packet center $({\bm r}_{cs}, {\bm k}_{cs})$ drift according to the semiclassical equations of motion \cite{Sundaram}
\begin{equation} \label{sc}
\arraycolsep 0.3 ex
\begin{array}{rl}
\displaystyle \hbar \dot {\bf k}_{cs} = & \displaystyle -e {\bf E} \\ [2ex]
\displaystyle \hbar \dot {\bf r}_{cs} = & \displaystyle \pd{\varepsilon_s}{{\bf k}_c} + e{\bf E} \times \bm{\mathcal F}_s,
\end{array}
\end{equation}
where $\bm{\mathcal F}_s = \bm{\nabla_k} \times \bm{\mathcal{R}}_{ss} $ represents the Berry, or geometrical curvature \cite{Sundaram}, with $\bm{\nabla_k}$ the gradient operator in momentum space. A careful analysis reveals that the Berry curvature also represents coherence between bands and must be taken into account in order to obtain agreement with approaches based directly on the Liouville equation. It is emphasized that the Berry curvature is not a result  of the particles being described by wave packets, which have a finite extent in real and momentum space. Rather, it emerges from the phase of the Bloch wave functions involved in the construction of the wave packets, implying that this formulation of semiclassical transport theory is a useful tool for capturing Berry-phase effects.

The electric field and disorder potential also give rise to a nonadiabatic mixing of the bands such that the basis states $\ket{{\bm k}s}$ become $\ket{\tilde{{\bm k}s}}$, given by
\begin{equation}
\label{uptb} 
\ket{\tilde{{\bm k}s}} = \ket{{\bm k}s} + \sum_{{\bm k}', s' \ne s}  [e\, {\bm E}\, \delta_{{\bm k}{\bm k}'} - ({\bm \nabla}U^{dis})_{{\bm k}{\bm k}'}] \cdot \frac{\bm{\mathcal{R}}_{s's}}{\varepsilon_{{\bm k}s} - \varepsilon_{{\bm k}s'}} \, \ket{{\bm k}s'}, 
\end{equation} 
where the $\ket{{\bm k}s}$ are the unperturbed eigenstates. The $\ket{\tilde{{\bm k}s}}$ also form a complete set. The distribution functions $f_s({\bm k}_{cs}, t)$ are made up of an equilibrium part and a part linear in the electric field, $f_s({\bm k}_{cs}, t) = f_{0s}({\bm k}_{cs}, t) + f_{Es}({\bm k}_{cs}, t)$, and the expectation values $\bkt{}_s$ in Eq.\ (\ref{Swpk}) also have terms of zeroth and linear orders in the electric field. Therefore to first order in the field the expectation value of operator $\hat{O}$ contains terms arising from the product of $f_{Es}({\bm k}_{cs}, t)$ with the zeroth order term in $\bkt{}_s$ as well as from the equilibrium $f_{0s}({\bm k}_{cs}, t)$ multiplied by the term in $\bkt{}_s$ linear in the electric field. 

\begin{figure}[tbp]
\centering \epsfig{file=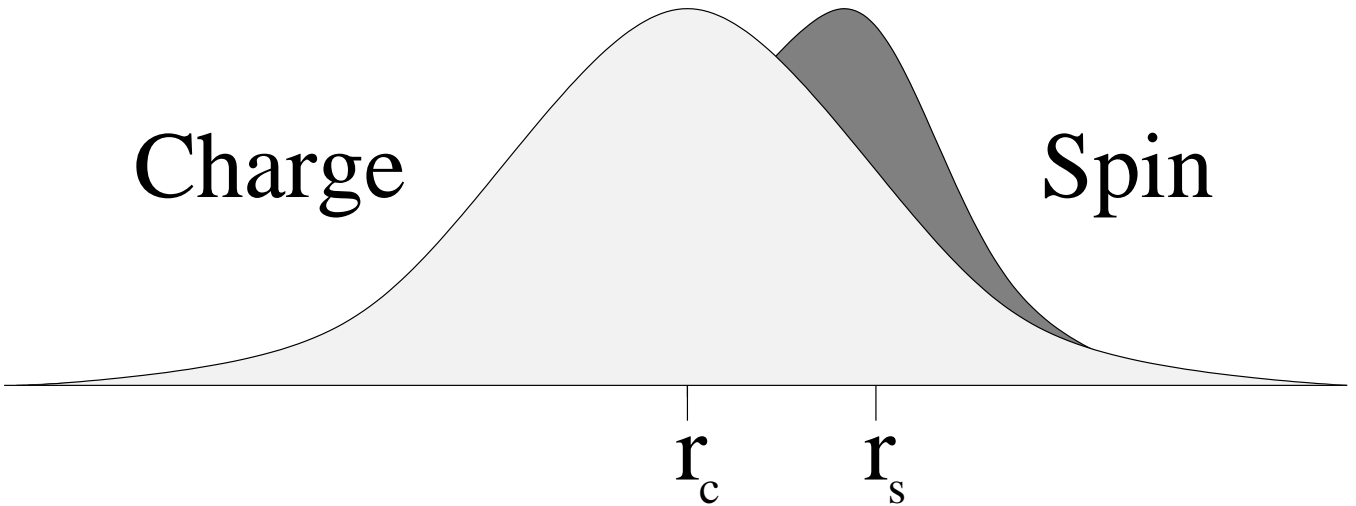, width=\columnwidth}
\caption{For a particle of finite extent the charge and spin
distributions in real space are in general do not coincide. The
same is true of the charge and spin distributions in reciprocal
space.} \label{wavepacket}
\end{figure}

The above exposition shows that interband coherence lost in passing from the density matrix to the Boltzmann approach is thus recovered in many ways. It is present in the Berry curvature, in the nonadiabatic correction to the wave functions, and in the interband scattering terms coupling the two distribution functions. The construction of a correct semiclassical theory that contains all the physics relevant to transport is seen to entail a sizable effort, particularly in the case of transport of non-conserved quantities such as spin.

\section{Band-structure spin densities and currents}

This work will now focus on the kinetic equation formalism and determine the spin densities and spin currents induced by external electric fields under a variety of circumstances. It is most straightforward to begin with effects originating in the band structure. The Hamiltonian of spin-1/2 electron systems typically contains a kinetic energy term and a spin-orbit coupling term, $H_{{\bm k}} = \frac{\hbar^2k^2}{2m^*} + H_{{\bm k}}^\mathrm{so}$, where $m^*$ is the electron effective mass. In spin-1/2 electron systems, band structure spin-orbit coupling can always be represented as a Zeeman-like interaction of the spin with a wave vector-dependent effective magnetic field ${\bm \Omega}_{\bm k}$, thus $H_{{\bm k}}^\mathrm{so} = (\hbar/2) \,{\bm \sigma}\cdot {\bm\Omega}_{\bm k}$. Common examples of effective fields are the Rashba spin-orbit interaction, \cite{ras84} which is often dominant in quantum wells with inversion asymmetry, and the Dresselhaus spin-orbit interaction, \cite{dre55} which is due to bulk inversion asymmetry. 

An electron spin at wave vector ${\bm k}$ precesses about the effective field ${\bm \Omega}_{\bm k}$ with frequency $\Omega_{\bm{k}}/\hbar \equiv |{\bm \Omega_{\bm{k}}}|/\hbar$ and is scattered to a different wave vector within a characteristic momentum scattering time $\tau$. Within the range  $\varepsilon_F\tau/\hbar \gg 1$, the relative magnitude of the spin precession frequency $\Omega_{\bm{k}}$ and inverse scattering time $1/\tau$ define three qualitatively different regimes. In the ballistic or clean regime no scattering occurs and the temperature tends to absolute zero, so that $\varepsilon_F\tau/\hbar \rightarrow \infty$ and $\Omega_{\bm{k}}\tau/\hbar \rightarrow \infty$. The weak scattering regime is characterized by fast spin precession and little momentum scattering due to, e.g., a slight increase in temperature, yielding $\varepsilon_F\tau/\hbar \gg \Omega_{\bm{k}} \, \tau /\hbar \gg 1$. In the strong momentum scattering regime $\varepsilon_F\tau/\hbar \gg 1 \gg \Omega_{\bm{k}} \, \tau/\hbar $. 

The kinetic equation ({\ref{eq:Fermi}}) has a scalar part which determines $n_{E{\bm k}}$
\begin{equation}\label{eq:nE}
\pd{n_{E{\bm k}}}{t} + \hat J_0 \, (n_{E {\bm{k}}}) = \frac{e{\bm
E}}{\hbar}\cdot\pd{n_{0{\bm k}}}{{\bm k}}.
\end{equation}
The solution of this equation is given by the well-known expression
\begin{equation}
n_{E{\bm k}} = \frac{e{\bm E}\tau_p}{\hbar}\cdot\pd{n_{0{\bm
k}}}{{\bm k}},
\end{equation}
in other words, $n_{E{\bm k}}$ describes the shift of the Fermi sphere in the presence of the electric field ${\bm E}$. The expression for the momentum relaxation time $\tau_p$ is a little different depending on the dimensionality of the system. Using $\gamma$ to denote the relative angle between ${\bm k}$ and ${\bm k}'$,
\begin{subequations}
\begin{eqnarray}\label{taup3d}
\frac{1}{\tau_p}& = & \frac{n_i mk}{2\pi \hbar^3} \int_0^\pi d\gamma \, |\mathcal{U}_{\bm{k}\bm{k}'}|^2 \, \sin\gamma \, (1 - \cos\gamma)  \,\,\,\,\,\, (3D) \\ [1ex]
\label{taup2d} && \frac{n_i m}{2\pi \hbar^3} \int_{0}^{2\pi}d\gamma \, |\mathcal{U}_{\bm{k}\bm{k}'}|^2 \, (1 - \cos\gamma) \,\,\,\,\,\, \,\,\,\,\,\,  \,\,\,\,\,\,\,(2D).
\end{eqnarray}
\end{subequations}

The spin-dependent part of the nonequilibrium correction to the density matrix $S_{E{\bm k}}$ is the spin density induced by ${\bm E}$. Its time evolution is governed by
$S_{E{\bm k}}$ is
\begin{equation}\label{eq:SBoltz} \pd{S_{E {\bm{k}}}}{t} + \frac{i}{\hbar}\, [H_{\bm k}, S_{E {\bm{k}}}] + \hat J_0 \, (S_{E {\bm{k}}}) = \frac{e{\bm
E}}{\hbar}\cdot\pd{S_{0{\bm k}}}{{\bm k}} - \hat J_b \, (n_{E {\bm{k}}}).
\end{equation}
Spin-dependent scattering gives rise to a renormalization of the driving term in the equation for $S_{E{\bm k}}$ with no analog in charge transport, see Eq.\ (\ref{eq:nE}).

Since an electron spin at wave vector ${\bm k}$ precesses about ${\bm \Omega}_{\bm k}$, the spin can be resolved into
components parallel and perpendicular to ${\bm \Omega}_{\bm k}$. In the course of spin precession the component of the spin parallel to
${\bm \Omega}_{\bm k}$ is conserved, while the perpendicular component is continually changing. It will prove useful in our
analysis to divide the spin distribution into a part representing conserved spin and a part representing precessing spin. The effective source term, which enters the RHS of Eq.\ (\ref{eq:SBoltz}), is divided into two parts, $(e{\bm E}/\hbar)\cdot\partial S_{0{\bm k}}/\partial{\bm k} - \hat J_b \,
(n_{E {\bm{k}}}) = \Sigma_{E{\bm k}\|} + \Sigma_{E{\bm k}\perp}$. $\Sigma_{E{\bm k}\|}$ commutes with $H^{so}_{\bm k}$ while $\Sigma_{E{\bm k}\perp}$ is \emph{orthogonal} to it and $\trace (\Sigma_{E{\bm k}\perp}\, H^\mathrm{so}_{\bm k}) = 0$. Projections onto and orthogonal to $H^\mathrm{so}_{\bm k}$ are most easily carried out by defining projectors $P_\|$ and $P_\perp$ by their actions on the basis matrices $\sigma_i$ as described in Ref.\ [\cite{dim07}]. $S_{E {\bm{k}}}$ is likewise divided into two terms: $S_{E {\bm{k}}\|}$, commuting with the spin-orbit Hamiltonian and $S_{E {\bm{k}}\perp}$, orthogonal to it. $S_{E {\bm{k}}\|}$ is the distribution of conserved spins while $S_{E {\bm{k}}\perp}$ is the distribution of precessing spins. Equation (\ref{eq:SBoltz}) is divided into separate equations for $S_{E {\bm{k}}\|}$ and $S_{E {\bm{k}}\perp}$
\begin{subequations}
\begin{eqnarray}
& & \pd{S_{E {\bm{k}}\|}}{t}  +  P_\| \hat{J}_0 \, (S_{E {\bm{k}}}) = \Sigma_{E{\bm k}\|}, \label{eq:Sigmaparallel}
\\ [1ex]
& & \pd{S_{E {\bm{k}}\perp}}{t}  + \frac{i}{\hbar}\, [H_{\bm k}, S_{E {\bm{k}}\perp}] + P_\perp \hat{J}_0 \, (S_{E {\bm{k}}}) = \Sigma_{E{\bm k}\perp} \label{eq:Sigmaperp}.
\hspace{3em}
\end{eqnarray}
\end{subequations}
The absence of the commutator $[H_{\bm k}, S_{E {\bm{k}}\|}] = 0$ in Eq.\ (\ref{eq:Sigmaparallel}) indicates the absence of spin precession, while the commutator $[H_{\bm k}, S_{E {\bm{k}}\perp}]$ in Eq.\ (\ref{eq:Sigmaperp}) represents spin precession. In order to solve Eqs.\ (\ref{eq:Sigmaparallel}) and (\ref{eq:Sigmaperp}) for arbitrary scattering, it is necessary to expand $S_{E {\bm{k}}\|}$ and $S_{E {\bm{k}}\perp}$ in $n_i$, as $S_{E {\bm{k}}\|} = S_{E {\bm{k}}\|}^{(-1)} + S_{E {\bm{k}}\|}^{(0)} + \mathcal{O} (n_i)$ and $S_{E {\bm{k}}\perp} = S_{E {\bm{k}}\perp}^{(0)} + \mathcal{O} (n_i)$. This is an expansion in the parameter $\hbar/(\Omega_{\bm{k}} \, \tau_p)$ and is most suited to systems in the weak scattering regime. The expansion of $S_{E {\bm{k}}\|}$ starts at order $-1$, a fact which can be understood by inspecting Eq.\ (\ref{eq:Sigmaparallel}). In the steady state the time derivative drops out, and the operator $\hat{J}_0$ is first order in $n_i$, while the right-hand side is independent of $n_i$. As a result, the expansion of the solution must start at order $-1$. Equation \ (\ref{eq:Sigmaperp}) for $S_{E {\bm{k}}\perp}$ tells us that, since $H_{\bm k}$ is independent of $n_i$ and the right hand side is also independent of $n_i$, the expansion of $S_{E {\bm{k}}\perp}$ must start at order zero. Equation (\ref{eq:Sigmaparallel}) can be solved iteratively for any scattering
\begin{equation} \label{eq:Sparminus1sol}
S_{E {\bm{k}}\|}^{(-1)} = \Sigma_{E{\bm k}\|} \tau_0 + P_\| \bigg(\frac{m^*}{2\pi \hbar^3} \int d\theta' |\mathcal{U}_{{\bm k}{\bm k}'}|^2 \, \Sigma_{E{\bm k}\|}\bigg) \tau_0^2 + \ldots 
\end{equation}
where $\tau_0 = n_i m^* \int d\theta' |\mathcal{U}_{{\bm k}{\bm k}'}|^2/(2\pi\hbar^3)$ is the quantum lifetime of the carriers. Above and henceforth integrals over wave vectors will be represented as two-dimensional, and $\theta'$ will refer to the polar angle of ${\bm k}'$. The extension to three dimensions is straightforward. The equations for higher orders in $n_i$ are easily deduced. However, the term of order $-1$ is by far the dominant one in the weak momentum scattering regime and is expected to be dominant over a wide range of strengths of the scattering potential.

It is evident that the steady state for conserved spins involves no spin precession, and that the correction $S_{E {\bm{k}}\|}$ depends explicitly on the nonequilibrium shift in the Fermi surface. In addition, scattering terms contain only the even function $ |\mathcal{U}_{{\bm k}{\bm k}'}|^2$. As a result, $S_{E {\bm{k}}\|}$ does not give rise to a spin current. Inspection of Eq.\ (\ref{eq:Sparminus1sol}) shows that integrals of the form $\int d\theta \, \hat{\mathcal{J}}^\sigma_{i} \, S_{E {\bm{k}}\|}$ contain an odd number of powers of ${\bm k}$ and are therefore zero. In the absence of impurity spin-orbit interactions, the distribution of conserved spins can give no spin current. It can, however, give rise to a nonequilibrium spin density since integrals of the form $\int d\theta \, \hat{s}^\sigma \, S_{E {\bm{k}}\|}$ contain an even number of powers of ${\bm k}$ and may be nonzero.

The leading term $S_{E {\bm{k}}\perp}^{(0)}$ is found to be
\begin{equation} \label{eq:Sperp0}
S_{E {\bm{k}}\perp}^{(0)} = \frac{\hbar}{2} \, \frac{{\bm \sigma} \cdot \hat{\bm\Omega}_{\bm k}\times [{\bm \Sigma}_{E {\bm{k}}\perp} - P_{\perp} \hat{J}_0 \, ({\bm S}_{E {\bm{k}}\|})]}{\Omega_{\bm k}},
\end{equation}
where we have written $\Sigma_{E {\bm{k}}\perp} = (1/2) \, {\bm \Sigma}_{E {\bm{k}}\perp}\cdot{\bm \sigma}$ and $S_{E {\bm{k}}\|} = (1/2) \, {\bm S}_{E {\bm{k}}\|}\cdot{\bm \sigma}$. The result expressed by Eq.\ (\ref{eq:Sperp0}) is valid for any elastic scattering. An argument similar to that given above shows that $S_{E {\bm{k}}\perp}^{(0)}$ cannot lead to a nonequilibrium spin density (although, as will be shown below, higher-order terms in $S_{E {\bm{k}}\perp}$ can contribute to the spin density). For, taking the expectation value of the spin operator, one arrives at integrals of the form $\int d\theta \, \hat{s}^\sigma \, S_{E {\bm{k}}\perp}^{(0)}$, which involve odd numbers of powers of ${\bm k}$ and are therefore zero. This term in the distribution of precessing spin does, however, give rise to nonzero spin currents, since integrals if the form $\int d\theta \, \hat{\mathcal{J}}^\sigma_{i} \, S_{E {\bm{k}}\perp}$ contain an even numbers of powers of ${\bm k}$ and may be nonzero. Consequently, in the absence of spin-orbit coupling in the scattering potential, nonequilibrium spin currents arise from spin precession.

The dominant contribution to the nonequilibrium spin density in an electric field exists because in the course of spin precession a component of each individual spin is preserved. For an electron with wave vector $\bm k$, this spin component is parallel to ${\bm \Omega}_{\bm k}$. In equilibrium the
average of these conserved components is zero. However, when an electric field is applied, the Fermi surface is shifted, and the average of the conserved spin components may be nonzero, as illustrated in Fig.~2. This intuitive physical argument explains why the nonequilibrium spin density $\propto \tau_p^{-1}$ and \emph{requires} scattering to balance the drift of the Fermi surface. It is interesting to note, also, that, although spin densities in electric fields require the presence of band structure spin-orbit interactions and therefore spin precession, the dominant contribution arises as a result of the absence of spin precession. Band structure spin currents on the other hand are associated with displacement of spins. The relation between spin currents and spin precession was made explicit by Sinova \textit{et~al.} \cite{sin04} Both $S_{E {\bm k} \|}$ and $S_{E {\bm k} \perp}$ are invariant under time-reversal. As a result, the tensor characterizing the response of spin currents to electric fields is invariant under time reversal, whereas the tensor characterizing the response of spin densities to electric fields changes sign under time reversal.

\begin{figure}[bt]
\centerline{\epsfig{file=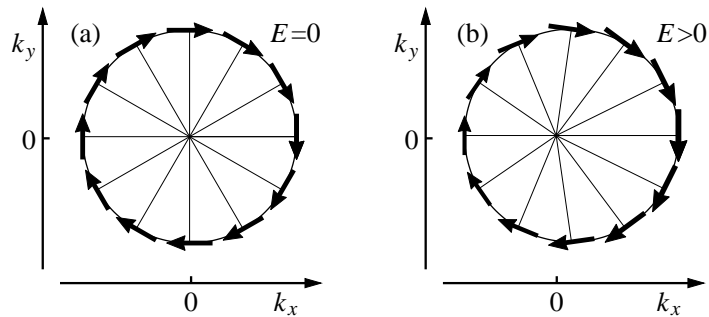, width=3.65in}}
\vspace*{8pt}
\caption{Effective field ${\bm \Omega_k}$ at the Fermi energy in the Rashba model (a) without ($E=0$) and (b) with ($E>0$) an external electric field.}
\label{fig:sia_e} 
\end{figure}

It will be noticed that the driving term in the equation for the nonequilibrium spin distribution $S_{E {\bm{k}}}$ is renormalized by the term $\hat J_s \, (n_{E {\bm{k}}})$, which accounts for spin-dependent scattering. In addition, Eq.\ (\ref{eq:Sigmaperp}) shows that scattering mixes
the distributions of conserved and precessing spins. When one spin at wave vector ${\bm k}$ and precessing about ${\bm \Omega}_{\bm k}$ is scattered to wave vector ${\bm k}'$ and precesses about ${\bm \Omega}_{{\bm k}'}$, its conserved component changes, a process which alters the distributions of conserved and precessing spin. Consequently, scattering processes in systems with spin-orbit interactions cause a renormalization of the driving term for the spin distribution as well as scattering between the conserved and precessing spin distributions. Contributions due to these two processes are contained in vertex corrections to spin-dependent quantities found in Green's functions formalisms. For short-range impurities it is easy to show \cite{dim07,RaimondiVertex} that the scattering correction $P_{\perp} \hat{J}_0 \, ({\bm S}_{E {\bm{k}}\|})$ to Eq.\ (\ref{eq:Sperp0}) depends on the steady state spin \emph{density}, and it becomes evident that the existence of a nonzero nonequilibrium spin density does affect the spin current. This fact was first pointed out by Raimondi \cite{RaimondiVertex} for the Rashba model.

The spin-Hall current was initially determined for spin-3/2 holes in GaAs \cite{mur03} and for an asymmetric quantum well \cite{sin04} in which the spin-orbit interaction is described by the Rashba model. The latter calculation yielded a spin-Hall conductivity $\sigma^z_{xy} = e/(8\pi)$, in which $\sigma^i_{jk}$ is understood as referring to spin component $i$ flowing in direction $j$ in response to an electric field applied along $k$. However, it was subsequently shown that a more careful treatment of disorder renormalizes this result to zero. In fact in two dimensions, for Hamiltonians linear in wave vector, the spin current vanishes for short-range impurities (Refs.~[\cite{ino04}-\cite{ll06}]), for small-angle scattering \cite{kha06,shy06} and in general for any elastic scattering. \cite{sug06} However, it does not vanish in a generalized Rashba model as was shown by Krotkov and Das Sarma. \cite{kro06} For spin-orbit Hamiltonians characterized solely by one angular Fourier component $N$ the spin current $\propto N$. \cite{shy06}

Ab initio calculation of band structure spin currents were also performed by Guo \textit{et al.} \cite{GuoSHE}  in semiconductors and by Yao \textit{et al.}\cite{YaoSign} in semiconductors and simple metals such as tungsten, platinum and gold. Spin transport in metals usually requires complex band structure calculations, which are typically done numerically. Yao \textit{et al.}\cite{YaoSign} found that the band structure spin-Hall effect in metals can be significantly larger than in semiconductors and that the spin-Hall conductivity can even undergo sign changes under certain circumstances.

\subsection{Spin adiabatically following a magnetic field}

The existence of a spin current in an electric field and its association with spin precession, can be understood from a straightforward argument due to Sinova.\cite{sin04} A spin precesses about an effective magnetic field ${\bm \Omega_k}$ which depends on ${\bm k}$. The electric field changes the wave vector ${\bm k}$ and in the process changes ${\bm \Omega_k}$, so the spin is now precessing about magnetic field that is changing adiabatically. In the adiabatic limit the spin follows the magnetic field. \cite{Sakurai} In particular, in 2D if the magnetic field is in
the plane of the spin, the spin never acquires a significant out-of plane component. However the spin in general acquires a small out-of-plane
component. Consider a generalized magnetic field ${\bm \Omega}$ and take the simple example in which the magnetic field $\parallel \hat{\bm x}$ and has a small $y$-component which increases linearly with time, such that $\Omega_y = \epsilon t$. The spin starts out along $x$, initially parallel to the field. In both cases, $s_x$ will be considered large compared to $s_y$, $s_z$ and in this section $\hbar = 1$. The adiabatic condition means that the magnitude of the magnetic field $\Omega$ changes little over one revolution of the electron spin about it. This means that, if we start at $t=0$ and consider a small time increment $\tau$, $\Omega (\tau) - \Omega (0) << \Omega (0)$.
\begin{equation}
\arraycolsep 0.3ex
\begin{array}{rl}
\Omega (\tau) = &\displaystyle \Omega (0) + \td{\Omega}{t}|_0 \tau \\[2.2ex]
\Omega (\tau) - \Omega (0) = &\displaystyle \td{\Omega}{t}|_0 \tau \rightarrow \td{\Omega}{t}|_0 \tau << \Omega (0).
\end{array}
\end{equation}
The adiabatic limit in this case is the limit of fast precession so one can choose as the small time $\tau$ the
(approximate) precession period around the magnetic field, $\tau = \frac{2\pi}{\Omega}$. The condition becomes
\begin{equation}
\arraycolsep 0.3ex
\begin{array}{rl}
\frac{2\pi}{\Omega}\td{\Omega}{t}  << \Omega.
\end{array}
\end{equation}
Consider the case in which a small damping term exists $\alpha (\td{{\bm s}}{t}\times {\bm s})$. The equations of motion for the three spin components are
\begin{equation}
\arraycolsep 0.3ex
\begin{array}{rl}
\td{s_x}{t} = &\displaystyle - s_z \Omega_y + \alpha (\td{s_y}{t} s_z - \td{s_z}{t} s_y)  \\ [2.2ex]
\td{s_y}{t} = &\displaystyle s_z \Omega_x + \alpha (\td{s_z}{t} s_x - \td{s_x}{t} s_z)   \\ [2.2ex]
\td{s_z}{t} = &\displaystyle s_x \Omega_y - s_y \Omega_x + \alpha (\td{s_x}{t} s_y - \td{s_y}{t} s_x).
\end{array}
\end{equation}
It is clear that all the changes in $s_x$ are at least of second order in small quantities, so $s_x$ can be treated as a constant. The solution is approximately
\begin{equation}
\arraycolsep 0.3ex
\begin{array}{rl}
\displaystyle s_y (t) = & \displaystyle \frac{\epsilon s_x}{\Omega_x^2} (\Omega_x t - e^{-t/\tau}\sin\Omega_xt) \\ [2ex]
\displaystyle s_z (t) = & \displaystyle \frac{\epsilon s_x}{\Omega^2} [1 - e^{-t/\tau}(\cos\Omega t - \frac{1}{\Omega \tau} \sin\Omega t)].
\end{array}
\end{equation}
where $\tau = \frac{1}{\alpha\Omega_x s_x}$. After the oscillations in $s_y$, $s_z$ are damped, what remains is
\begin{equation}
\arraycolsep 0.3ex
\begin{array}{rl}
\displaystyle s_y (t >> \tau) = &\displaystyle \frac{\epsilon s_x}{\Omega_x} t \\[2.2ex]
\displaystyle s_z (t >> \tau) = &\displaystyle \frac{\epsilon s_x}{\Omega^2} .
\end{array}
\end{equation}
The spin, which was originally in the plane, acquires a small steady-state out-of-plane component. This component depends on $s_x$, and $s_x$ has the opposite sign on different sides of the Fermi surface. As a result, $s_z$ up and down spins travel in opposite directions and a spin-Hall current is established. This derivation, illustrated by Fig.~3, is another way to clarify the role of spin precession in steady-state band structure spin currents.

\begin{figure}[bt]
\centerline{ \psfig{file=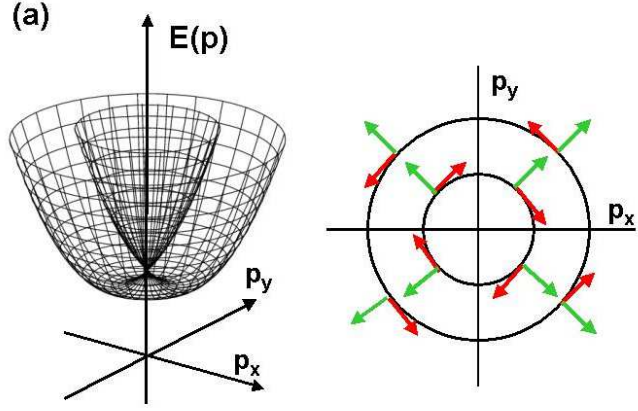, width=3.65in}}
\vspace*{8pt}
\caption{Spin splitting of energy spectrum in the Rashba model. The green arrows represent the direction of the momentum and the red arrows represent the direction of the spin.}
\label{fig3a} 
\end{figure}

\begin{figure}[bt]
\centerline{ \psfig{file=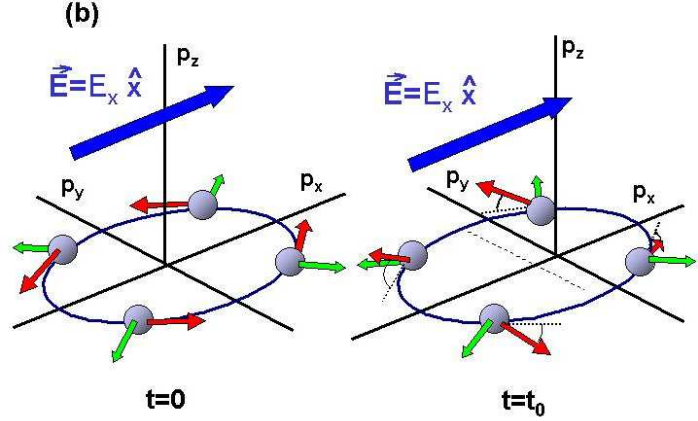, width=3.65in}}
\vspace*{8pt}
\caption{An electric field displaces the Fermi surface and the electron spins are tilted up for $p_y > 0$ and down for $p_y < 0$.}
\label{fig3b} 
\end{figure}

\section{Extrinsic spin densities and currents}

Until now the scattering potential has been treated as a scalar. However, in general the spin-orbit interaction makes a contribution to the disorder potential. The disorder potential including the scalar and spin-orbit parts takes the form 
\begin{equation}
\arraycolsep 0.3ex
\begin{array}{rl}
\displaystyle U_{{\bm k}{\bm k}'} = & \displaystyle [1 + i \, \lambda \, {\bm \sigma} \cdot ({\bm k} \times {\bm k}')] \, \frac{\mathcal{U}_{{\bm k}{\bm k}'}}{V}.
\end{array}
\end{equation}
with $\mathcal{U}_{{\bm k}{\bm k}'}$ the matrix element of the \emph{scalar} part of the potential. The second term comes from spin-orbit coupling in $U_{{\bm k}{\bm k}'}$ and $\lambda$ is a material-specific constant. Configurational averages of terms of the form $U^{imp}_{{\bm k}{\bm k}'}U^{imp}_{{\bm k}'{\bm k''}}U^{imp}_{{\bm k}''{\bm k}}$ are also linear in $n_i$ but are rather lengthy to be displayed explicitly.

The spin-orbit interaction also produces a correction\cite{Nozieres} to the position operator $\hat{\bm r} = \hat{\bm r}_{ord} + 2 \lambda \, {\bm \sigma} \times {\bm k}$ where $\hat{\bm r}_{ord}$ is the ordinary position operator. Since the external electric field enters through the Hamiltonian $H_E = e{\bm E} \cdot \hat{\bm r}$ the correction to $\hat{\bm r}$ gives a \textit{side jump} term $H_E^j = 2 e \lambda {\bm \sigma} \cdot {\bm k} \times {\bm E}$ and this side jump term makes an additional contribution\cite{Nozieres} to the velocity operator ${\bm v}_E^j = - \frac{2e \lambda}{\hbar} \, {\bm \sigma} \times {\bm E}$. In the kinetic equation extra \textit{spin-dependent driving term} $ - (i/\hbar)\, [H_E^j, f_{0{\bm k}}]$ must be taken into account due to the change in the position operator. This term is nonzero if the equilibrium density matrix $f_{0{\bm k}}$ is spin-dependent. The side jump mechanism is typically classified as extrinsic although its contributions are manifold. These mechanisms were studied extensively in the anomalous Hall effect \cite{Lutt,Smit,Berger,Nozieres} and recently their role in steady-state spin densities and currents have been highlighted in semiconductors\cite{eng05,ono06,EwelinaC,ChengWu} and metals.\cite{Shche} 

The solution for $n_{E{\bm k}}$ is found as above and the spin-dependent scattering term$\hat{J}_s $ acts on $n_{E{\bm k}}$ and produces an additional source terms for $S_{E{\bm k}}$, such that the RHS of Eq.\ (\ref{eq:SBoltz}) becomes
\begin{equation}\label{eq:S}
\begin{array}{rl}
\displaystyle \Sigma_{E{\bm k}} - \hat{J}_s (n_{E{\bm k}}) - \frac{i}{\hbar} \, [H_E^j, S_{0{\bm k}}] - \hat{J}_j (f_{0{\bm k}}).
\end{array}
\end{equation}
Skew scattering emerges in the second Born approximation as a third-order term in the potential $\mathcal{U}$. Substituting for $n_{E{\bm k}}$ and introducing the angles $\gamma_1 = \theta' - \theta$, $\gamma_2 = \theta'' - \theta'$, $\gamma_3 = \theta - \theta''$ and the solid angles $\omega'$, $\omega''$, 
\begin{equation}
\arraycolsep 0.3ex
\begin{array}{rl}
\displaystyle \hat{J}_s(n_{\bm k}) = & \displaystyle  - \frac{6\pi^2 \lambda n_i e \tau_p m^{*2} k^{2d-2}}{\hbar^6} \, \delta(k - k_F) \, {\bm \sigma} \cdot \hat{\bm k} \times \bm{\mathcal I} \\ [2ex] 
\displaystyle \bm{\mathcal I} = & \displaystyle \int \frac{d\omega'} {(2\pi)^d}  \int \frac{d\omega''}{(2\pi)^d} \, \mathcal{U}(\gamma_1) \mathcal{U}(\gamma_2) \mathcal{U}(\gamma_3) \, (\hat{\bm k}') \, {\bm E}\cdot (\hat{\bm k}' - \hat{\bm k}'').
\end{array}
\end{equation}
The integral over solid angles $\bm{\mathcal I}$ is independent of $\hat{\bm k}$. In two dimensions the integrand is expanded in Fourier harmonics and the integration is straightforward. In three dimensions the integrand is expanded in Legendre polynomials of $\gamma_1$, $\gamma_2$ and $\gamma_3$, and the integral over the two solid angles of $P_l(\gamma_1)P_m(\gamma_2)P_n(\gamma_3)$ is independent of $\hat{\bm k}$. The modification of the position operator produces an additional \textit{side jump} scattering term
\begin{equation} 
- \hat{J}_j (f_{0{\bm k}}) = - \frac{ H_E^j \, \delta(\varepsilon_{{\bm k}} - \varepsilon_F)}{\tau_p}.
\end{equation}
This term is related quite literally to transverse jumps undergone by a carrier during scattering \cite{Nozieres} and its form reflects the conservation of momentum during such a side jump scattering event. \cite{Nozieres,EwelinaC} 

The intrinsic source $\Sigma_{E{\bm k}}$ was discussed in the previous section. The extrinsic source is henceforth denoted by $\mathcal{T}_{E{\bm k}} = - \frac{i}{\hbar} \, [H_E^j, S_{0{\bm k}}] - \hat{J}_j (f_{0{\bm k}}) - \hat{J}_s (n_{E{\bm k}})$. If the band structure spin-orbit interactions are zero one finds easily
\begin{equation}
S^{ext}_{E{\bm k}} = H_E^j \, \delta(\varepsilon_{{\bm k}} - \varepsilon_F) - \hat{J}_s (n_{E{\bm k}})\, \tau_p,
\end{equation}
where the superscript $ext$ indicates that only extrinsic contributions are considered. This expression averages to zero over directions in momentum space and does not give a spin density. It does however  contribute to the spin current, giving a spin-Hall conductivity $\propto n_i^{-1}$ due to skew scattering and a spin-Hall conductivity $ne\lambda$ independent of $n_i$ due to side jump. These terms were discussed by Engel \textit{et al.} \cite{eng05} and by Hankiewicz and Vignale \cite{EwelinaC}.

If band structure spin-orbit interactions are nonzero then, as was pointed out by several groups,\cite{TseIntExt,EwelinaP,ChengWu,HuExt} spin precession is crucial in establishing the steady state. To illustrate this consider once again the decomposition $S^{ext}_{E{\bm k}} = S^{ext}_{E{\bm k}\parallel} + S^{ext}_{E{\bm k}\perp}$ into linearly independent components. In two-dimensional systems grown along (001) the entire extrinsic source term is orthogonal to $H_{so}$, therefore $\mathcal{T}_{E{\bm k}\|} = 0$ and there is no term in $S^{ext}_{E{\bm k}}$ that is $\propto n_i^{-1}$. The solution is 
\begin{equation}\label{Sp}
\displaystyle S^{ext}_{E{\bm k}\perp} = \frac{{\bm \sigma}\cdot\hat{\bm \Omega}_{\bm k} \times \bm {\mathcal T}_{E{\bm k}}}{2\Omega_{\bm k}}.
\end{equation}
The contributions to $S^{ext}_{E{\bm k}\perp}$ due to skew scattering and side jump contain ${\bm k}$ to an even power, therefore skew scattering and side jump do not contribute to the spin current in the presence of spin precession (they can be restored by an external magnetic field $\bm{B} \parallel \hat{\bm z}$ as found in Ref.\ [\cite{EwelinaP}].) The only contribution to the spin current comes from the spin-dependent driving term. The spin-Hall conductivity originating from this term is $ne\lambda$ regardless of the form of the band structure spin-orbit interaction. Skew scattering in the presence of band structure spin-orbit interactions does give a steady-state spin density, which was pointed out in [\cite{ChengWu}], and a similar spin density arises from side jump.

It is therefore seen that skew scattering and side jump contribute very differently to the steady-state density matrix when band structure spin-orbit interactions are present. This can be understood from the following argument. In order to get a sizable spin current from skew scattering, spins must be conserved as they travel. Scattering processes produce a separation between spin-up and spin-down, while precession tries to destroy the orientation of the spins. Thus once the spins are scattered it is the conserved spin fraction that carries the spin current. Evidently the same argument applies for side jumps undergone during scattering.

\section{Definitions of the spin current and boundary conditions}

The spin current is defined in an intuitive manner by $\hat{\mathcal{J}}^\sigma_{i} = (1/2)\, \{ s^\sigma, v^i \}$, and this definition is used by the majority of researchers working on this topic. Nevertheless, in the presence of band structure spin-orbit interactions this spin current is not conserved. The equation of continuity satisfied by the spin density and current was shown by Shi \textit{et al.} to be different \cite{shi06} from the usual equation of continuity for the charge density and current. A source term exists in this equation which reflects spin non-conservation. As a result, Shi \textit{et al.} as well as Bryksin and Kleinert \cite{Bryksin} have proposed an alternative definition of the spin current according to which $\hat{\mathcal{J}}^\sigma_{i} = d/dt\, (\hat r_i \hat s^\sigma)$. The equation of continuity satisfied by this current still contains a source term, but this source term vanishes in many commonly used models. This definition was used in Refs.\ [\cite{zar06,sug06}]. Recently, Tokatly \cite{Tokatly} has argued that the intuitive definition of the spin current represents a dissipative current which is conjugate to an effective SU(2) electric field. 

In order to obtain the spin accumulation the kinetic equation needs to be supplemented by boundary conditions. This was done by a number of groups.  \cite{TserkovnyakAcc,BleiAcc,UsajAcc,MalshukovAcc,NomuraAcc,TseAcc,GalitskiAcc,ada05} Unfortunately, it was shown that the form of the spin accumulation depends on the theoretical model of the boundary. \cite{TseAcc,GalitskiAcc} The nontrivial physics associated with boundary conditions and the implications of various formulations of boundary conditions for spin transport were discussed by Bleibaum \cite{BleiAcc} and Tserkovnyak \textit{et al.}\cite{TserkovnyakAcc} An interesting argument due to Adagideli and Bauer \cite{ada05} points out that, in systems in which the band structure spin current is expected to vanish in the bulk (i.e. for Rashba and Dresselhaus spin-orbit coupling), the situation is not the same near the edges and near metal contacts, where a finite spin current exists which can be detected.

\section{Crystal symmetry}

Spin densities and spin currents in a crystal are closely tied to the symmetry of the underlying lattice. An analysis relating the response tensor to the symmetry of the underlying crystal lattice has been enlightening in the context of nonequilibrium spin densities excited by an electric field. If the response of the spin density $\bm{s}$ to an electric field $\bm{E}$ is given by $s^\sigma = Q^{\sigma}_j E_j$, nonzero components for the material-specific spin density response tensor $Q^{\sigma}_j$ are permitted only in gyrotropic crystals \cite{vor79}. For spin transport such an analysis was performed in Ref.[\cite{dim07}], determining the components of the spin-current response tensor allowed by symmetry in an electric
field and providing systematic proof that spin currents in response to an electric field can be much more complex than the spin-Hall
effect \cite{nik04}. This result is completely general and is not sensitive to the definition of the spin current or to whether the
electric field is constant or time-dependent. 

The spin current operator can be defined as $\hat{\mathcal{J}}^\sigma_{i} = {\textstyle\frac{1}{2}} (\hat{s}^\sigma \hat{v}_i + \hat{v}_i \hat{s}^\sigma )$ or $\hat{\mathcal{J}}^\sigma_{i} = d/dt\, (\hat r_i \hat s^\sigma)$. From a symmetry point of view these two definitions are equivalent. The spin current $\hat{\vekc{J}}$ is a second rank tensor that can be decomposed into a pseudoscalar part, an antisymmetric (spin-Hall) part, and a symmetric part. The pseudoscalar part is $\trace (\hat{\vekc{J}}) = {\textstyle\frac{1}{3}} \, \hat{\bm s} \cdot \hat{\bm v}$ and represents a spin flowing in the direction in which it is oriented. The symmetric and antisymmetric parts are given, respectively, by ${\textstyle\frac{1}{2}} (\hat{s}^\sigma \hat{v}_i \pm \hat{v}_\sigma \hat{s}^i )$. The pseudoscalar and symmetric parts will be referred to as \emph{non-spin-Hall} currents. Under the full orthogonal group only the antisymmetric (spin-Hall) components of $\hat{\vekc{J}}$ are allowed, indicating that these components are always permitted by symmetry.

In general, the spin current response of a crystal to an electric field $\bm{E}$ is characterized by a material tensor $\bm{T}$ defined by $\mathcal{J}^\sigma_{i} = T^\sigma_{ij} E_j$. For the 32 crystallographic point groups the symmetry analysis \cite{Bir} for the tensor $\bm{T}$ is established by means of standard compatibility relations \cite{Koster}. One is particularly interested in those groups in which non-spin-Hall components may be present. The pseudoscalar part of the spin current is only allowed by 13 point groups, while the symmetric part is allowed in all systems except those with point groups $O$, $T_d$ (zinc blende), or $O_h$ (diamond). Lower symmetries, allowing non-spin-Hall currents, are characteristic of systems of reduced dimensionality. In such systems it was shown that non-spin Hall currents exist.\cite{dim07} For a quantum well grown along (113), the effective magnetic field corresponding to the Rashba spin-orbit interaction has a component in the $\hat{\bm z}$-direction as well, and Ref.\ [\cite{dim07}] showed that the spin conductivities $\sigma^x_{xx}$, $\sigma^x_{yy}$, $\sigma^z_{yx}$ and $\sigma^z_{xy}$ are nonzero.

Skew scattering and side jump contributions to the spin current in the presence of band structure spin-orbit interactions can also be restored by growing the structure along a direction that is not one of the main crystal axes. The contribution to the spin density matrix due to extrinsic mechanisms has two parts,
\begin{equation}
\arraycolsep 0.3ex
\begin{array}{rl}
\displaystyle S^{ext}_{E{\bm k}\parallel} = & \displaystyle \frac{1}{2} \, \bigg(\frac{{\mathcal T}_{E{\bm k}z}\Omega_{{\bm k}z}}{\Omega_{\bm k}^2}\bigg) \, {\bm \sigma}\cdot{\bm \Omega_k} \, \tau
\end{array}
\end{equation}
and $S^{ext}_{E{\bm k}\perp}$, which is given by Eq.\ (\ref{Sp}). Taking again a quantum well grown along (113), $S^{ext}_{E{\bm k}\parallel}$ gives no spin density but gives a spin current from skew scattering and side jump. Skew scattering and side jump give the same nonzero components of the spin conductivity tensor, $\sigma^x_{xx}$, $\sigma^x_{yy}$, $\sigma^z_{yx}$ and $\sigma^z_{xy}$. 

\section{Experimental situation}

Spin densities in semiconductors were first predicted by Ivchenko \cite{ivc78} and observed the following year in tellurium by Vorob'ev. \cite{vor79} Recent experiments\cite{kato04b,GanichevJMMM,sil04} have also reported the observation of a steady-state spin density in semiconductors. Experimentally the spin density can be found by Kerr rotation, in which a beam of linearly-polarized light is sent into the sample and its polarization vector rotates by an angle that is proportional to the spin polarization. 

In so far as measuring a spin current, the situation is more complicated. Initially the approach adopted was to generate a spin current which would flow to the edges of the sample where it would produce a spin accumulation. This spin density at the edge of the sample could then be detected by Kerr rotation or an equivalent technique. J.~Wunderlich \textit{et al.} \cite{wun05} used photoluminescence to detect a spin accumulation due to a spin-Hall current in a two dimensional hole gas, which is believed to be due to spin-orbit interactions in the band structure. Kato \textit{et al.} \cite{kato04} used Kerr rotation to detect a spin accumulation due to a spin-Hall current  in $n$-GaAs, in which the spin current is believed to be due to extrinsic mechanisms \cite{eng05}. A similar experiment was carried out shortly thereafter by Sih \textit{et al.} \cite{sih05}. Stern \textit{et al.} \cite{ste06} used the same technique to detect a spin-Hall current in ZnSe at room temperature. Following this work, Sih \textit{et al.} \cite{AwschGen} performed an experiment designed to demonstrate explicitly that the observed edge spin accumulation was due to the spin-Hall effect, as opposed to edge effects associated with the electric field. The group manufactured samples with a series of transverse channels, such that the spin-Hall current generated by the electric field was allowed to drift into regions where the electric field was effectively zero. Since a spin accumulation was still measured at the edge, this showed unambiguously that the effect was due to the spin current. Further experiments by Stern \textit{et al.}\cite{AwschDriftDiff} imaged the spatial distribution of the spin accumulation generated by the spin-Hall effect as well as its behavior in a magnetic field. Chang \textit{et al.}\cite{Chang} also reported, using photoluminescence, a spin accumulation due to the spin-Hall effect in InGaN/GaN superlattices.

A second type of experiments relies on an effect referred to as the inverse spin-Hall effect. Briefly, a spin current in turn generates a charge current, which can be detected by conventional means. This was shown by Hirsch \cite{hir99} and its extension to a Landauer-Buttiker type multi-terminal nonlocal measurement was presented by Hankiewicz \textit{et al}.\cite{EwelinaLan} This technique was used by Valenzuela and Tinkham \cite{val06} to detect a charge current as a result of the spin current in aluminium, and by Saitoh \textit{et al.} \cite{SaitohInvSHE} in platinum. Kimura \textit{et al.} \cite{Kimura} observed the spin-Hall effect in platinum at room temperature and their findings were explained theoretically by Guo \textit{et al.}, \cite{GuoPt} who demonstrated that the effect is due to band structure spin-orbit interactions. Vila \textit{et al.} also obtained results for platinum nanowires \cite{Vila} while Seki \textit{et al.} \cite{Seki} reported a giant spin Hall effect in FePt/Au devices. The group used a multi-terminal device with a gold cross and FePt acting as a spin injector. The spin-Hall resistance was measured to be 2.9 m$\Omega$ and is attributed to the large skew scattering in gold, being thus extrinsic in nature. Weng \textit{et al.} \cite{Singapore} carried out similar experiments on platinum, aluminium and gold and obtained results in agreement with those found to date. It should be noted that  Cui \textit{et al.}\cite{cui06} also observed an electrical current induced by an optically-generated spin current.
	
\section{Future Directions}

Whereas the scientific community working on steady-state spin densities and currents appears to be in agreement that spin currents exist and are experimentally measurable, a number of questions remain to be addressed in the future. For example, the relative magnitude of band structure spin currents and spin currents due to extrinsic mechanisms such as skew scattering and side jump remains to be determined for a general band structure spin-orbit interaction. In addition, the fact that no unique definition of the spin current exists causes difficulties in the comparison of experimental data with theoretical predictions. This ambiguity is exacerbated by the fact that different definitions of the spin current give results that often differ by a sign. \cite{shi06,sug06} Thanks to the non-conservation of spin, the relationship between spin current and spin accumulation at the boundary remains to be clarified. It appears that what happens at the boundary is sensitive to the type of boundary conditions assumed.\cite{TseAcc,GalitskiAcc} Thus so far as quantitative interpretation of experimental data is concerned, theory still has some way to go. 

On the experimental side, despite tremendous progress, the community is still searching for a reliable way to \emph{measure}, as opposed to \emph{detect}, spin currents directly. A possible new experimental avenue relies on magnetoresistance caused by an edge spin accumulation, proposed by Dyakonov. \cite{DyakMagRes} A spin current produces a spin accumulation near the sample edges, which in turn causes the sample resistance to decrease by a small amount, whereas an external magnetic field can destroy the edge spin polarization and yield a positive magnetoresistance. An alternative path was followed by Wang \textit{et al.},\cite{wang06} who started from the Dirac equation and obtained in the weakly relativistic limit a set of Maxwell equations in the presence of spin-orbit interactions. Although the spin current does not appear explicitly, it is contained in the Maxwell equations and the authors demonstrate that the relativistic conservation laws imply that the spin current yields an electrical polarization, which could be detected directly. Similarly, a set of Maxwell-like equations for spin was formulated by Bernevig \textit{et al.}\cite{BernevigMaxwell} These methods remains to be attempted experimentally.

Practically, the question of what to do with the electron spin once it has been generated or transported to the edge of the sample remains. The revolutionary electronic device that harnesses spin currents for a practical purpose remains to be made, and the challenge of its design confronts experimentalists and theorists alike.

The research at Argonne National Laboratory was supported by the US Department of Energy, Office of Science, Office of Basic Energy Sciences, under Contract No. DE-AC02-06CH11357.

\end{document}